\documentclass[prl,twocolumn,showpacs,preprintnumbers,amsmath,amssymb]{revtex4}

\usepackage{graphicx}
\usepackage{dcolumn}
\usepackage{bm}
\usepackage{color}

\begin{document}
\newcommand{\js}[1]
{
\textcolor{red}{{#1}}
}

\preprint{APS/123-QED}

\title{Surface bubble nucleation phase space}
\author{James R. T. Seddon$^1$}
\email{j.r.t.seddon@utwente.nl}
\author{E. Stefan Kooij$^2$}
\author{Bene Poelsema$^2$}
\author{Harold J. W. Zandvliet$^{2}$}
\author{Detlef Lohse$^1$}
\affiliation{$^1$Physics of Fluids,  $^2$Solid State Physics,  MESA+ Institute for Nanotechnology, University of Twente, P.O. Box 217, 7500 AE Enschede, The Netherlands}
\begin{abstract}
Recent research has revealed several different techniques for nanoscopic gas nucleation on submerged surfaces, with findings seemingly
 in contradiction with each other.
In response to this, we have  systematically
investigated the occurrence of surface nanobubbles on a hydrophobised silicon substrate for various different liquid temperatures and gas concentrations, which we controlled independently.  We found that nanobubbles occupy a distinct region of this phase space, occurring for gas concentrations of approximately $100-110\,\%$.
Below the nanobubble phase we did not detect any gaseous formations on the substrate, whereas \textit{micropancakes} (micron wide, nanometer high gaseous domains) were found at higher temperatures and gas concentrations.  We moreover find that supersaturation of dissolved gases is \textit{not} a requirement for nucleation of bubbles.
\end{abstract}
\maketitle

Bubbles normally rapidly dissolve unless the system is out of equilibrium.  \textit{Surface nanobubbles} are an exception \cite{parker1994,tyrrell2001,tyrrell2002,borkent2007,zhang2008,hampton2010}.  These gaseous bubbles are trapped at the solid-liquid interface, with typical dimensions of $\sim 100\,\mathrm{nm}$ width and $10-20\,\mathrm{nm}$ height \cite{borkent2009}.  Their corresponding radii of curvature are usually less than $1\,\mathrm{\mu m}$ \cite{borkent2010}, thus the prediction from the  standard diffusion model would be complete dissolution in $\sim 1\,\mathrm{\mu s}$ in degassed water.  However, surface nanobubbles have been found to persist for \textit{at least} $5\,\mathrm{days}$ \cite{ducker2009}, some $11$ orders of magnitude longer than the simple expectation. Understanding this apparent stability to diffusion has been the prime research question on nanobubbles ever since their discovery.

One possible explanation for this discrepancy is that nanobubbles become coated with
 diffusion-limiting contaminants \cite{ducker2009}, which is disheartening but possibly true.  An alternative
 explanation is that gas is diffusing out of the nanobubbles, but  that
this diffusive outflux is stably balanced by an influx of gas at the contact line \cite{brenner2008}.  Whatever the stabilizing factor is, nanobubbles remain on the solid-liquid interface until the experiment is stopped and the system dried, or until the liquid evaporates.  Hence, nanobubbles can be considered stable.

We  now turn our attention to  nanobubble nucleation.  Several important observations already exist: Electrolysis is an obvious creation mechanism \cite{zhang2006b,yang2009}.
Also local supersaturation of dissolved gases in the liquid
is considered to be a possibility
 for high-density nucleation.  Indeed,
 substrate heating \cite{zhang2004,yang2007} and alcohol-water exchange \cite{zhang2004,yang2008} are
 prominent techniques, too.
  The claim is that these latter two methods result in the system being pushed far from thermodynamic equilibrium, through the introduction of  both temperature (direct heating, or exothermic reaction) and gas concentration (supersaturation due to heating) variations.  However, it is unknown whether one, or both, of these variations is the most important, and further research is urgently required to solve this mystery.

In this Letter we directly investigate this specific point by \textit{independently} controlling liquid temperature and dissolved gas concentration during deposition.  Our observations are four-fold: (i) Liquid temperature and dissolved gas concentration variations can both  lead to the creation of nanobubbles, (ii) supersaturation is not an essential ingredient for nucleation, (iii) nanobubbles are created in a distinct region of our phase space, and (iv) pushing the system further above equilibrium leads to the preferential creation of a different gaseous domain, namely micropancakes \cite{zhang2007,seddon2010a}.

We first describe our experimental method.  The substrate was a silicon wafer that had been hydrophobised with perfluorodecyltrichlorosilane (PFDTS), following the guidelines of Ref.~\cite{seddon2010a}.  It was then mounted on a temperature controlled sample plate (331 temperature controller, Lakeshore, USA) and a purpose-built atomic force microscope (AFM) liquid cell was firmly mounted on top.

For the liquid,  pure water was prepared using a Simplicity 185 system (Millipore, France).  The water flask was placed on top of a hotplate with feedback control (Ika, Germany), and the temperature and oxygen content of the water were measured using a PSt3 oximeter (PreSens, Germany).  The oximeter gave a very accurate ($\lesssim 0.5\,\%$) reading of oxygen concentration, and we recorded the concentration in both milligrams per litre and percentage saturation.  Furthermore, we used the oxygen concentration to estimate air saturation, bearing in mind that diffusion coefficients of nitrogen and oxygen are approximately the same ($D_{N_2} \approx 2.0\times 10^{-9} \,\mathrm{m^2/s}$; $D_{O_2} \approx 2.4\times 10^{-9} \,\mathrm{m^2/s}$) \cite{crc}.

We imaged the substrate with an Agilent 5100 AFM in tapping mode.  The cantilevers were hydrophilic, Au-back-coated Si$_3$N$_4$ Veeco NPG probes (radius of curvature $30\,\mathrm{nm}$, full tip cone angle $35\,\mathrm{^o}$), with resonance frequencies in liquid of $\omega_0^{liq} \approx 15-25\,\mathrm{kHz}$.  In the current experiments we operated the AFM at a frequency typically $0.2\,\mathrm{kHz}$ lower than resonance, with a set point of $90\,\%$.

The experimental method was as follows: (i) Set the substrate temperature to the desired value; (ii) set the  temperature of the hotplate on which the water flask is placed to a certain value; (iii) use  slow heating/cooling to adjust the gas concentration; (iv) when the oximeter and temperature readings were at the required values, extract liquid from the water flask with a glass and metal syringe and deposit onto the substrate (this procedure lasted no longer than $5-6\,\mathrm{s}$);  (v) scan the substrate (which took $\sim10\,\mathrm{mins}$ per scan of $2\,\mathrm{\mu m}\times 2\,\mathrm{\mu m}$).

This whole procedure
 has allowed us to investigate directly the effects of both gas concentration and liquid temperature
{\it independently} and \textit{immediately after deposition} on nanobubble formation.

We now report our findings, beginning with gas-saturated water in equilibrium with the substrate, i.e. the substrate and liquid were set to the same temperature (i.e. $T_{liq}=T_{sub}$).  This is significantly different to previous temperature studies, whereby the liquid parameters were unknown due to use of the ethanol-water exchange \cite{zhang2004}, by thermally ramping the substrate temperature \cite{zhang2005,yang2007}, or by rapidly heating the liquid prior to deposition \cite{yang2007}.

\begin{figure}
\begin{center}
\includegraphics[width=8cm]{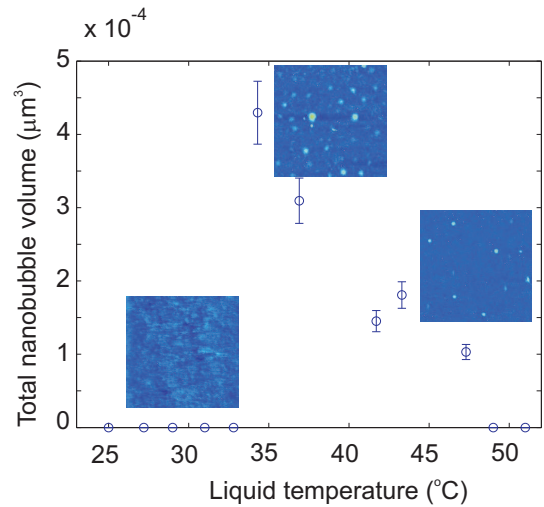}
\end{center}
\caption{Total volume of nanobubbles on a $2\,\mathrm{\mu m}\, \times \,2\,\mathrm{\mu m}$ area of PFDTS, after deposition, as a function of temperature.  The liquid and substrate are held at the same temperature throughout each measurement, and the liquid is $100\,\%$ saturated with gas prior to deposition.  The insets are typical images. \label{fig:100pc}}
\end{figure}

Our first observation is that the supersaturation of gases dissolved in the liquid is \textit{not} a prerequisite for nanobubble formation.   To demonstrate this we plot the total volume of all nanobubbles on a $2\,\mathrm{\mu m}\, \times \,2\,\mathrm{\mu m}$ area, as a function of liquid temperature, in Fig. \ref{fig:100pc}, where we re-stress that the liquid was $100\,\%$ saturated with gas (note that the absolute gas concentration monotonically decreases with increasing liquid temperature).

No nanobubbles were nucleated when the temperature was below $\sim33\,\mathrm{^oC}$, whereas the volume of nanobubbles vastly increased at $\approx34\,\mathrm{^oC}$.  We currently have no explanation for this
abrupt
switching transition, but we note that temperatures elevated above room temperature are usually required for nanobubble nucleation.
Repeating the measurements at higher temperatures led to a decrease in the total nanobubble volume, as expected: The amount of gas dissolved in saturated liquid decreases with increasing temperature, so less gas was trapped at the interface when there was less gas available.

We now turn our attention to the effects of changing liquid temperature and the concentration of gases dissolved in the water, on a room temperature substrate.  In this respect, we set $T_{sub}$ to $21\,\mathrm{^oC}$ for the remainder of our measurements, and used liquid temperature and gas concentration \textit{prior to deposition} as the control parameters, i.e. the dissolved gas was in equilibrium with the water at the chosen temperature, but this was however  different from the substrate temperature.  Due to the small volume of liquid in our AFM liquid cell,  the water rapidly cooled to $T_{sub}$ (within $\tau \sim D^2/\kappa \approx (2\,\mathrm{mm})^2/1.4\times10^{-7}m^2/s = 10-20\,\mathrm{s}$, where $D$ is the thickness of the fluid layer and $\kappa$ is the thermal diffusivity) prior to the first scan, so our measurements provide snapshots of the effect of liquid temperature and gas concentration at the time of deposition.

\begin{figure}
\begin{center}
\includegraphics[width=8cm]{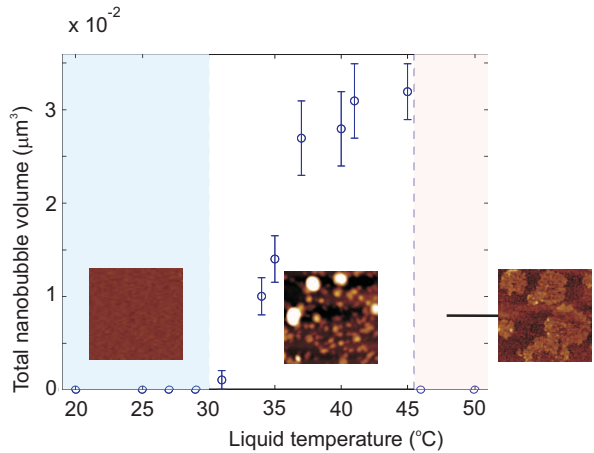}
\end{center}
\caption{Total nanobubble volume on a $2\,\mathrm{\mu m}\times2\,\mathrm{\mu m}$ area as a function of liquid temperature, on a room temperature substrate.  For each measurement the liquid was $100\,\%$ saturated with air prior to deposition.  For temperatures less than $\sim 30\,\mathrm{^oC}$ no gaseous domains formed; nanobubbles were formed for temperatures in the range $30\,\mathrm{^oC}\lesssim T \lesssim 45\,\mathrm{^oC}$; whilst micropancakes were formed for temperatures above $\sim 45\,\mathrm{^oC}$.  Shading is given as a guide to the eye, and the insets are typical images from each of the three regions. \label{fig:vol}}
\end{figure}

In Fig. \ref{fig:vol} we plot the results of nanobubble formation in $100\,\%$ saturated water as a function of liquid temperature when following this procedure.  Three distinct regions exist: (i) No nucleation for $T_{liq}\lesssim 30\,\mathrm{^oC}$; (ii) nanobubble nucleation for $30\,\mathrm{^oC}\lesssim T_{liq} \lesssim 45\,\mathrm{^oC}$, with an increasing amount of gas trapped with increasing temperature; (iii) \textit{micropancake} nucleation  (gaseous domains with typical diameters of microns and heights of $\sim 1\,\mathrm{nm}$) \cite{zhang2007,seddon2010a} for $T_{liq} \gtrsim 45\,\mathrm{^oC}$ (examples of each are shown in the insets).

Remarkably,
 the temperature dependence of nanobubble volume is the {\it opposite}
 to that found with the $100\,\%$ saturated liquid used in Fig.\
 \ref{fig:100pc}, but is in agreement with far-from-equilibrium measurements of Refs. \cite{zhang2004,zhang2005,yang2007}.  Our explanation for our finding
 is that although less gas is available as the liquid temperature is increased prior to deposition,
a larger gas concentration gradient is set up in the direction of the substrate at impact, so we may expect locally higher gas concentrations at the solid/liquid interface.

In Fig.\ \ref{fig:vol}, there is also a new regime, in which {\it  micropancakes} nucleate.  The exact structure of micropancakes is currently unknown, but they are most probably adsorbates.  Then, the simplest explanation for the transition from surface nanobubble to micropancake formation is the trapping of air molecules at the solid interface during wetting by the condensing vapour.  Note that we did not observe the \textit{coexistence} of nanobubbles and micropancakes, as seen on  highly-orientated-pyrolytic-graphite  (HOPG) by Zhang \textit{et al.} \cite{zhang2006d}.
An explanation may be that, if micropancakes are indeed
adsorbates,  we would expect them to be much stabler on HOPG due to the underlying crystal structure.

\begin{figure}
\begin{center}
\includegraphics[width=8cm]{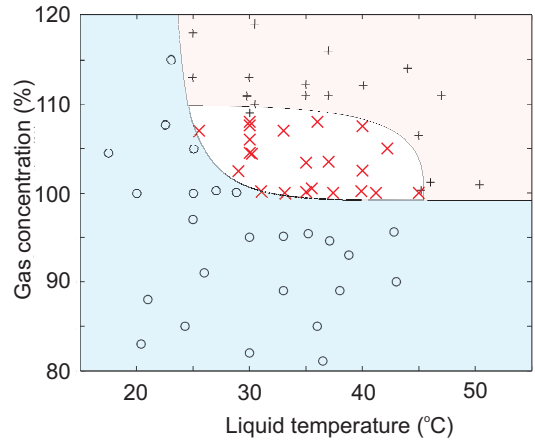}
\end{center}
\caption{Regions of nanobubble nucleation in percentage gas concentration -- liquid temperature phase space, for a room temperature substrate.  Nanobubbles were only found for the system parameters marked with a cross.  Below this region (on both axes) no gaseous domains were formed (circles), whilst \textit{micropancakes} preferentially formed far above this region (pluses).  Shading and solid lines are given as guides to the eye.\label{fig:percentage}}
\end{figure}

We now extend the measurements above to investigate the effect of different concentrations of gas and liquid temperatures, i.e. the liquid is no longer in equilibrium with the dissolved gas, \textit{and} is at a different temperature to the substrate.   We plot our results in the phase space gas concentration vs.\ liquid temperature,
 see Fig.\ \ref{fig:percentage}, in which  the symbols correspond to no nucleation (circles), nanobubbles (crosses), or micropancakes (pluses) \cite{comment_100pc}.
Nanobubbles are only formed in a small region of
 this phase space, in-between the extensive regions of zero nucleation and micropancake formation.  In total, we explored the phase space from $\sim 65-120\,\%$ gas concentration and for liquid temperatures in the range $\sim 17-50\,\mathrm{^oC}$, but we only observed the formation of nanobubbles in the single region shown in Fig. \ref{fig:percentage}.  This explains the lack of reproducibility in the field to date: A $1-2\,\mathrm{K}$ variation in temperature or a $1-2\,\%$ variation in gas concentration can simply push you out of the nanobubble-nucleating regime.  These key parameters must be controlled.

\begin{figure}
\begin{center}
\includegraphics[width=8cm]{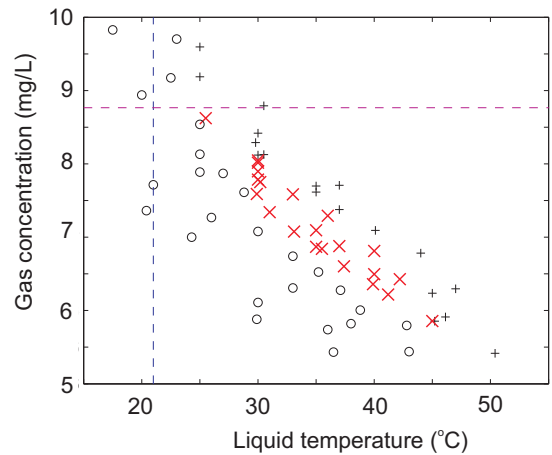}
\end{center}
\caption{Absolute gas concentration vs. liquid temperature for the data point in the phase space of Fig. \ref{fig:percentage}, with nanobubble nucleation (x), micropancake nucleation (+), and  no gaseous formation (o).  The vertical dashed line at $21\,\mathrm{^oC}$ represents the temperature of the substrate, whilst the horizontal dashed line represents the mg/L gas concentration that \textit{would be} in equilibrium with water at $21\,\mathrm{^oC}$.\label{fig:mgL}}
\end{figure}
Our data can also be displayed in a different way.  \textit{Percentage} gas concentration is itself a temperature dependent quantity.  Thus we replot Fig. \ref{fig:percentage}, but now with the gas concentration in absolute units (milligrams per liter), in Fig. \ref{fig:mgL}.    The motivation
 for replotting in this way is to demonstrate that nanobubble nucleation \textit{requires} that the liquid \textit{must} be undersaturated with respect to a room temperature substrate: For all the data in Fig. \ref{fig:vol} the liquid  cools to the substrate temperature and becomes undersaturated in $10-20\,\mathrm{s}$.

In order to shed some light on this observation, we return to the possible stabilizing factor for nanobubbles: Gaseous influx at the contact line \cite{brenner2008}.  Originally, it was assumed that supersaturation was essential to provide excess gas at the solid-liquid interface \cite{dammer2006}, but more recently it has been shown that this effect occurs with little dependence on dissolved gas concentration \cite{sendner2009}.  Thus, even though cooling the liquid should lead to local undersaturation, we still expect an excess of gas at the solid-liquid interface.  The fact that we do not see nanobubbles beyond the micropancake regime is indicative that the micropancake regime maybe far more extensive, perhaps with nanobubbles forming as a result of micropancake decay as the system becomes close to the transition of zero nucleation.

In conclusion, we have demonstrated that the supersaturation of gases dissolved in water is \textit{not} a requirement for the nucleation of nanobubbles.  Furthermore, for a given substrate temperature,  nanobubble nucleation during deposition is a strong function of both the liquid temperature \textit{and} the concentration of gas dissolved in the liquid.  Mapping out a phase space on these axes has allowed us to find a distinct region of this phase space in which nanobubbles readily nucleate.  Below the transition line, no gaseous domains form on the substrate, whereas far above the transition line we found a further transition separating the nucleation of nanobubbles and micropancakes.

Micropancakes only formed in the systems where the liquid had to cool to the substrate temperature after deposition, suggesting that micropancakes are adsorbates that condense out of the liquid phase.

From our findings it is now possible to nucleate either nanobubbles or micropancakes in a systematic way \textit{without} the risk of cross-contamination from, for example, the alcohol-water exchange.  Furthermore, it is now clear that ``safe'' zones exist in phase space that guarantee zero nucleation of either nanobubbles or micropancakes, thus ensuring that industrial processes for which nanobubbles would be detrimental (such as immersion lithography) can operate without cause for concern.

Finally, we return to the `standard' technique of creating surface nanobubbles, where it has  been previously thought that supersaturation was a key ingredient.  We performed oximetry tests on $200\,\mathrm{mL}$ of water  that had come straight out of our purifier.  The air concentration was $\sim 60\,\%$ and it took approximately $50\,\mathrm{hrs}$ to reach saturation.  We then performed oximetry tests on the water during the ethanol-water exchange.  The exothermic reaction led to a $4-5\,\mathrm{K}$ increase in temperature and a $\approx 25\,\%$ increase in
relative
gas concentration (of which only about half was accounted for by the increase in temperature, whilst the rest was probably due to the differing saturation levels of the two liquids at equilibrium).  Hence, simply extracting water from a purifier, waiting for a few hours, then using the alcohol-water exchange method, does not guarantee supersaturation.  In fact, in light of our present results, the water in previous experiments was most likely undersaturated.

We acknowledge preliminary measurements by Sezer Caynak.  The research leading to these results has received funding from the European Community's Seventh Framework Programme (\textit{FP7/2007-2013}) under \textit{grant agreement} number 235873, and from the Foundation for Fundamental Research on Matter (FOM), which is sponsored by the Netherlands Organization for Scientific Research (NWO).

\end{document}